\RequirePackage{fix-cm}
\documentclass[pdftex,twocolumn,epjc3]{svjour3}          
\RequirePackage[T1]{fontenc}
\smartqed  
\RequirePackage{graphicx}
\RequirePackage{mathptmx}      
\RequirePackage{flushend}
\RequirePackage[numbers,sort&compress]{natbib}
\RequirePackage[colorlinks,citecolor=blue,urlcolor=blue,linkcolor=blue]{hyperref}

\def\mevc{\ensuremath{\mathrm{MeV/}c}}
\def\gevc{\ensuremath{\mathrm{GeV/}c}}
\def\mev2c{\ensuremath{\mathrm{MeV/}c^2}}
\def\gev2c{\ensuremath{\mathrm{GeV/}c^2}}
\def\to{\ensuremath{\,\rightarrow\,}}

\def\pbarp{\ensuremath{{\bar{p}p}}}
\def\PiPi{\ensuremath{{\pi\pi}}}
\def\omegapi0{\ensuremath{\omega\pi^0}}

\def\PiPiEta{\ensuremath{{\pi^0\pi^0\eta}}}
\def\PiEtaEta{\ensuremath{{\pi^0\eta\eta}}}
\def\KpKmPi0{\ensuremath{{K^+K^-\pi^0}}}
\def\pbarpToPi0Pi0Eta{\ensuremath{{\bar{p}p}\,\rightarrow\,\pi^0\pi^0\eta}}
\def\pbarpToPiEtaEta{\ensuremath{{\bar{p}p}\,\rightarrow\,\pi^0\eta\eta}}
\def\pbarpToKpKmPi0{\ensuremath{{\bar{p}p}\,\rightarrow\,K^+K^-\pi^0}}

\def\pbarpToOmegaPi0{\ensuremath{{\bar{p}p}\,\rightarrow\,\omega\pi^0}}
\def\pbarpToPhiPi0{\ensuremath{{\bar{p}p}\,\rightarrow\,\phi(1020)\pi^0}}

\def\OmegaToPi0Gamma{\ensuremath{\omega\,\rightarrow\,\pi^0\gamma}}
\def\OmegaToPipPimPi0{\ensuremath{\omega\,\rightarrow\,\pi^+\pi^-\pi^0}}
\def\PipPimPi0{\ensuremath{\pi^+\pi^-\pi^0}}

\def\MassPi1{\ensuremath{(1623 \, \pm \, 47 \, ^{+24}_{-75})\, \mathrm{MeV/}c^2}}
\def\WoUnitsMassPi1{\ensuremath{1623 \, \pm \, 47 \, ^{+24}_{-75}}}
\def\WidthPi1{\ensuremath{(455 \, \pm 88 \, ^{+144}_{-175})\, \mathrm{MeV}}}
\def\WoUnitsWidthPi1{\ensuremath{455 \, \pm 88 \, ^{+144}_{-175}}}

\usepackage{multirow}

\usepackage{adjustbox}

\usepackage{amsmath}

\journalname{Eur. Phys. J. C}
\begin{document}

\title{Investigation of the Lightest Hybrid Meson Candidate with a
Coupled-Channel Analysis of \pbarp\,-, $\pi^- p$\,- and
\PiPi\,-Data}

\author{
  B.~Kopf\thanksref{addr1} \and
  M.~Albrecht\thanksref{addr1} \and
  H.~Koch\thanksref{addr1} \and
  M.~K{\"u}{\ss}ner\thanksref{addr1} \and
  J.~Pychy\thanksref{addr1} \and
 X.~Qin\thanksref{addr1, e1} \and
  U.~Wiedner\thanksref{addr1}
  }

\thankstext{e1}{~Now at Shandong University, 266237 Qingdao, China}

 \institute{%
   ~Ruhr-Universit\"at Bochum, 44801 Bochum, Germany\label{addr1}
  }
\date{Received: date / Accepted: date}

\maketitle
\begin{abstract}
  A sophisticated coupled-channel analysis is presented  that combines different
   processes: the channels \linebreak
   \PiPiEta, \PiEtaEta\ and \KpKmPi0 from \pbarp\
   annihilations, the P- and D-wave amplitudes of the $\pi\eta$ and
   $\pi\eta^\prime$ systems 
   produced in  $\pi^- p$ scattering, and data
from \PiPi-scatt\-ering reactions.
Hence our analysis combines the data sets used in two
independent previous analyses published by the Crystal Barrel
experiment and by the JPAC group. Based on the new insights
from these studies, this paper aims at a better understanding
of the spin-exotic $\pi_1$ resonances in the light-meson sector. By utilizing the K-matrix approach
and realizing the ana\-ly\-ti\-city
via Chew-Mandelstam functions the amplitude of the
  spin-exotic wave can be well described by a single
  $\pi_1$ pole for both systems,
$\pi\eta$ and
$\pi\eta^\prime$. The mass and the width of the $\pi_1$-pole  
are measured
to be \MassPi1\ and  \WidthPi1.

\end{abstract}

\section{Introduction}
The picture of $\pi_1$ resonances with spin-exotic quantum numbers
$I^G(J^{PC})$ = $1^-(1^{-+})$ in the light-meson
sector is poorly understood and the experimental indications of
various resonances are controversially discussed.
Lattice QCD calculations~\cite{Lacock:1996ny,Bernard:1997ib,Dudek:2013yja,Woss:2020ayi} and phenomenological QCD
studies~\cite{PhysRevD.65.025012,Szczepaniak:2006nx} predict only one state  at a mass of
2 GeV$/c^2$ or slightly below. Experimentally, three different
resonances with $I^G(J^{PC})$ = $1^-(1^{-+})$ quantum numbers have been
reported. The lightest one, the $\pi_1(1400)$, has only been seen in
the $\pi \eta$ decay mode by several
experiments~\cite{Alde:1988bv, Aoyagi:1993kn, Thompson:1997bs, Abele:1998gn, Abele:1999tf, Salvini:2004gz, Adams:2006sa}.
In contrast, for the $\pi_1(1600)$ no coupling to $\pi \eta$ has been found,
but it has been
observed in several other channels, namely $\pi \eta^\prime$, $\rho \pi$, $f_1(1285) \pi$ and
$b_1(1235) \pi$~\cite{Adams:1998ff, Alekseev:2009aa, Ivanov:2001rv, Kuhn:2004en, Lu:2004yn, Akhunzyanov:2018lqa}.
The third state which has
 the poorest evidence, and is thus listed in the Review
   of Particle Physics (RPP) as a further state, is the $\pi_1(2015)$ seen by the BNL E852 experiment decaying into
$f_1(1285) \pi$ and $b_1(1235) \pi$~\cite{Kuhn:2004en, Lu:2004yn}. A
weak point in various of these previous analyses is the
extraction of the resonance
parameters using Breit-Wigner parameterizations. The
outcome of an
analysis performed by the JPAC group likely sheds more
light on the understanding of the lightest $\pi_1$
states~\cite{Rodas:2018owy}. Utilizing the N/D method to model the reaction process, it turned out that the two candidates for
a spin-exotic state, $\pi_1(1400)$ and $\pi_1(1600)$, that
  are listed in the RPP, can be
described by only one pole with a separate coupling to $\pi\eta$ and $\pi\eta^\prime$.\\
The Crystal Barrel Collaboration observed
a significant $\pi_1$ contribution in \pbarp\ annihilations in
flight for the first time with a coupling to $\pi\eta$ in the reaction
\pbarpToPi0Pi0Eta~\cite{Albrecht:2019ssa} using a coupled-channel
analysis. In this paper, the analysis has been extended by considering not only the channels \pbarpToPi0Pi0Eta, \PiEtaEta\ and \KpKmPi0 at
a beam momentum of 900 \mevc\, and data from 11 different $\pi\pi$-scatt\-ering channels but also the P- and D-waves in the
$\pi\eta$ and $\pi\eta^\prime$ systems measured at
COMPASS~\cite{Adolph:2014rpp,Adolph:2014rppCorrigendum}.
The dynamics is treated
slightly differently compared to~\cite{Rodas:2018owy}. The K-matrix approach was used by taking into account the analyticity
with Chew-Mandelstam functions~\cite{PhysRevD.91.054008}.

\section{ Partial-Wave Analysis}
\label{sec:PWA}
The partial-wave analysis has been performed with the software package PAWIAN ({\bf PA}rtial {\bf W}ave {\bf I}nteractive {\bf AN}alysis
Software)~\cite{Kopf:2014wwa} and with the same algorithms as
described in~\cite{Albrecht:2019ssa}.

\paragraph*{Description of the \pbarp\ channels:}
~ We analyzed data of the three \pbarp\ annihilation channels \PiPiEta, \PiEtaEta\ and \KpKmPi0 at
a beam momentum of 900 \mevc\ measured with the Crystal Barrel detector at LEAR. The full description of the reconstruction and event selection can
be found in~\cite{Albrecht:2019ssa}.\\
The complete reaction chain starting from the \pbarp\ initial state
down to the final-state particles is fitted. The
  description for the angular part of the amplitudes is based
on the helicity formalism. The amplitudes are further expanded into the LS-scheme
which guaranties that also the orbital angular momentum
dependent barrier factors for the production and the decay
 are properly taken into account. In addition to  $\pi_1\pi^0$,  the sub-channels $f_0\eta$, $f_2\eta$, $a_0\pi^0$ and
$a_2\pi^0$ are contributing to \pbarpToPi0Pi0Eta. The contributing
isovector states $a_0$ and $a_2$ exhibit a similar decay pattern
as the $\pi_1$-wave amplitude with a strong
coupling to the $\pi^0\eta$ system. Therefore, it is not straightforward to properly disentangle these waves from each other.
In this case, the simultaneous fit of the channels \PiEtaEta\ and \KpKmPi0 helps considerably. It ensures a strong control on the production
of $a_0\pi^0$ and $a_2\pi^0$ by directly sharing the relevant
amplitudes between the two channels \PiPiEta\ and \KpKmPi0. Apart from
the two isolated resonances $\phi(1020)$ and $K^*(892)^\pm$,
 which
are described by Breit-Wigner functions, the K-matrix formalism with
P-vector approach is used for the
dynamics~\cite{Albrecht:2019ssa,AITCHISON1972417,PWAinKMatChungBrose}.\\
For each partial wave with defined quantum numbers $I^G(J^P)$, the
mass-dependent amplitude $F^{p}_i(s)$ is parametrized as follows:
\begin{equation}
  \label{equ:fvector}
F^{p}_i(s) \; = \; \sum_j (I \; + \; K(s) \; C(s))_{ij}^{-1} \cdot P^p_j(s),
\end{equation}
where $i$ and $j$ represent the two-body decay channels like $\pi\pi$,
$\pi\eta$ or $K^+K^-$ and $s$ is the invariant mass squared of the
respective two-body sub-channel. The analyticity is taken
  into account by using the Chew-Mandelstam function $C(s)$ \cite{PhysRev.119.467,PhysRevD.91.054008}.
$P^p_{j}(s)$ represents one element of the P-vector taking into
account the production process $p$:
\begin{equation}
  \label{equ:production}
P^p_j(s) =  \sum_{\alpha} \Big( \frac{\beta^p_{\alpha_p} \; g^{\text{bare}}_{\alpha_j} }{{m^{\text{bare}}_\alpha}^2
  -s} +\sum_k  c_{kj} \cdot s^k \Big) \cdot   B^l(q_j,q_{\alpha_j}),
\end{equation}
where $\beta^p_{\alpha_p}$ is the complex parameter representing the
strength of the produced resonance $\alpha$. $g^{\text{bare}}_{\alpha_j}$ and
$m^{\text{bare}}_\alpha$ are the bare parameters for the coupling strength to
the channel $j$ and for the mass of the resonance
$\alpha$. $B^l(q_j,q_{\alpha_j})$ denotes the Blatt-Weisskopf barrier
factor of the decay channel $j$ with the orbital angular momentum $l$, the breakup momentum $q_j$ and the
resonance breakup momentum $q_{\alpha_j}$. This factor
  was chosen such that the centrifugal barrier is explicitly included
  and  is normalized at $q_{\alpha_j}$ \cite[chap. 49]{Zyla:2020zbs}. It is worth
mentioning that the production barrier factors are already taken into
account separately in the production amplitude as described
in~\cite{Albrecht:2019ssa}. The
$s$-dependent polynomial terms of the order $k$ with
the parameters $c_{kj}$ describe background contributions for the
production.\\
The elements of the K-matrix for the two-body scattering process with orbital angular momentum
$l$ are given by
\begin{eqnarray}
  \label{equ:kmatrix}
K_{ij}(s)  = \sum_{\alpha} B^l(q_i,q_{\alpha_i}) \cdot \Big(
  \frac{g^{\text{bare}}_{\alpha_i} \; g^{\text{bare}}_{\alpha_j} }{{m_\alpha^{\text{bare}}}^2
  -s} +  \tilde{c}_{ij} \Big)  \cdot B^l(q_j,q_{\alpha_j}),
\end{eqnarray}
where $i$ and $j$ represent the input and output channels, respectively. The parameters $\tilde{c}_{ij}$
stand for the constant terms of the
background contributions, which are allowed to be added to the K-matrix without
violating unitarity. In addition to the form given in
Eq.~(\ref{equ:kmatrix}), the K-matrix elements for the description of
the $f_0$-wave amplitude are multiplied with an Adler
zero term $(s - s_0) / s_{\text{norm}}$ as outlined in
detail in~\cite{Albrecht:2019ssa}.\\
Also the K-matrix descriptions of the $f_0$-, $f_2$-, $\rho$-, $a_0$- and
$(K\pi)_S$-waves are the same as outlined in \cite{Albrecht:2019ssa}.
A slightly different K-matrix for the $\pi_1$- and $a_2$-wave is employed for the following reasons:
\begin{itemize}
\item the $\pi_1$-wave amplitude consists of one K-matrix pole and the two channels $\pi\eta$ and $\pi\eta^\prime$. Constant
      background terms for the K-matrix and the P-vector for the \pbarp\ channel have been used. Due to the fact that the $t$-channel exchange
      plays an important role for the  $\pi^-p$-scattering process, a
      first-order polynomial is needed for the relevant background terms of
      these P-vector elements.
\item the K-matrix of the $a_2$-wave is parametrized by two poles,
  $a_2(1320)$ and $a_2 (1700)$, and by the three channels
      $\pi \eta$, $\pi \eta^\prime$ and $K \bar{K}$. Constant
      background terms for the K-ma\-trix are used. No background terms for the P-vector are need\-ed for the \pbarp\ channels while also here
      first-order polynomial background terms are required for the
      production in $\pi^-p$. This is due to the fact that the COMPASS data are simultaneously
      fitted, which contain not only the $\pi\eta$ but also the
      $\pi\eta^{\prime}$ channel as discussed below.
    \end{itemize}
 
\paragraph*{Description of the \PiPi-scattering data:} 
~The mass-dependent terms of each partial wave describing the \PiPi-scattering
reactions are parametrized by the T-matrix
\begin{equation}
T(s) = (I \, + \,  K(s) \; C(s))^{-1} \; K(s),
\end{equation}
where $s$ is the total energy squared of the $\pi\pi$ system. For elastic channels, phase and
inelasticity are compared with the data. For inelastic channels,
such as
$\pi\pi \rightarrow \eta \eta$, the moduli squared of the T-matrix are
taken. As in~\cite{Albrecht:2019ssa}, the following data for the $I$ = 0\,
$S$- and $D$-wave and the $I$ = 1\,$P$-wave for energies below $\sqrt{s}
\, <\,1.9\,$ GeV were included in the analysis:
\begin{itemize}
\item the phases and inelasticities of the reaction $\pi\pi \rightarrow \pi\pi$ for the $I=0$\ $S$- and $D$-wave and for the $I=1$ $P$-wave,
\item the intensities of the inelastic channels $\pi\pi \rightarrow K\bar{K}$ and $\pi\pi \rightarrow \eta \eta $ for the $I=0$\ $S$- and
$D$-wave and
\item the intensity of the inelastic channel $\pi\pi \rightarrow \eta \eta^\prime$ for the $I=0$\ $S$-wave.
\end{itemize}

\paragraph*{Description of the COMPASS data:}
~The COMPASS data are taken from the mass-independent analysis of $\pi p \rightarrow \eta^{(\prime)} \pi p$ with a 191 \gevc\
pion beam integrated over the transferred momentum squared
$-t$ between 0.1 and 1.0 $\mathrm{GeV}^2$~\cite{Adolph:2014rpp,Adolph:2014rppCorrigendum}. Only the
intensities of the P- and D-wave and their relative phases for the two channels $\pi\eta$ and $\pi\eta^\prime$ are considered.
The choice of these waves for the coupled-channel analysis ensures strong constraints in particular for the channel \PiPiEta, where large
contributions of the $\pi_1$- and $a_2$-waves are found.\\
The description of the reaction $\pi p \rightarrow R p \rightarrow (\eta^{(\prime)} \pi) p$
with $R$ being the produced $\pi_1$ or  $a_2$ partial wave is approximated via
the exchange of a Pomeron with $J^P$\,=\,1$^-$ and an effective
transferred momentum squared  of
$t_{\text{eff}}  = -0.1\, \mathrm{GeV}^2$. The intensities
$I^{\pi p \rightarrow R p}_{\pi \eta^{(\prime)}}$ of
the COMPASS data for the P- and D-partial-wave are
described by:
\begin{equation}
I^{\pi p \rightarrow R p}_{\pi \eta^{(\prime)}} \; = \; p^{2J-2}_{\pi \eta^{(\prime)}} \cdot q_{\pi \eta^{(\prime)}} \cdot \big| F^{\pi p \rightarrow R p}_{\pi \eta^{(\prime)}} \big|^2,
\end{equation}
where $J$ stands for the spin associated to the relevant wave, $q_{\pi \eta^{(\prime)}}$
for the  $\pi \eta^{(\prime)}$ breakup momentum representing the behavior
of the phase space for the decay and $p_{\pi \eta^{(\prime)}}$ is the $\pi$ beam momentum in the $\pi
\eta^{(\prime)}$ rest frame, where
\begin{equation}
    p_{\pi\eta^{(\prime)}}=
    \frac{\sqrt{\lambda(s, m^2_\pi, t_{\text{eff}})}} {2\,\sqrt{s}}
\end{equation}
is the production breakup momentum with $\lambda$ being the K\"all\'en
triangle function. $F^{\pi p \rightarrow R p}_{\pi \eta^{(\prime)}}$ is taken as
given in Eq.~(\ref{equ:fvector}) and the production barrier
factor is taken into account by $p^{2J-2}_{\pi
  \eta^{(\prime)}}$ according to
~\cite{Jackura:2019ccv,Rodas:2018owy}.\\
The relative phases of the $\pi_1$- and $a_2$-wave amplitudes for the channels $\pi \eta$
and $\pi \eta^\prime$ are modeled as defined in
Eq.~(\ref{equ:fvector}) by subtracting the relevant
$F$-vectors according to \linebreak
$ F^{\pi p \rightarrow a_2 p}_{\pi\eta^{(\prime)}} - F^{\pi p \rightarrow \pi_1 p}_{\pi
    \eta^{(\prime)}}$.

\paragraph*{Fits to all data:}
~A combined minimization function is used for the fit, in which all
data sets are taken into account. For the \pbarp\ data,
 for each event the full information of the multi-dimensional phase
 space is used. The other data sets are provided as data points with uncertainties.
The construction of the
complete negative log-likelihood function to be minimized is performed
in analogy
to~\cite{Albrecht:2019ssa} and described in detail therein.\\
Different hypotheses have been tested by systematically
  add\-ing and removing the potentially contributing resonances. For the selection of the
  best fit hypothesis the Bayesian and the
  Akaike information criterion have been used in the same way as
  in~\cite{Albrecht:2019ssa}.
The significantly best fit result was achieved with the same
contributing resonances as in our previous paper~\cite{Albrecht:2019ssa}.
All tested
  hypotheses with the obtained information criteria
  are summarized in Tab.~1 of the supplemental material.

\section{Fit Result}
\begin{figure}[b]
\hspace{-3mm}\includegraphics[width=0.5\textwidth]{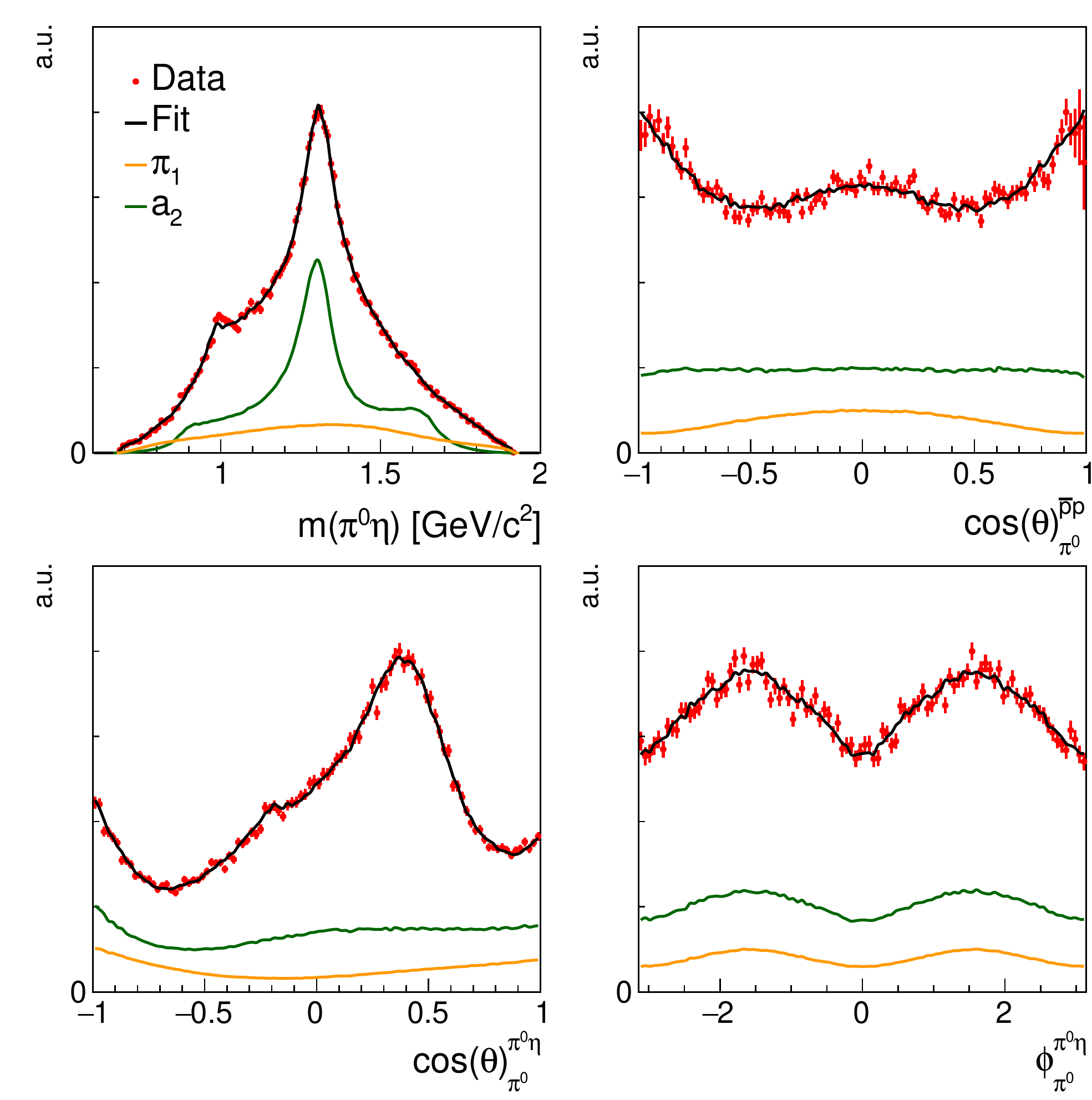}\\[-4ex]
   \caption{\label{fig:pipieta} Invariant mass distribution (upper
     left) and selected angular distributions for the production
     (upper right) and decay (lower left and right) of the \PiPiEta\
     channel in the \pbarp\ data. The red markers with error bars show
     the efficiency-corrected data with two entries per
     event, while the black curve
     represents our best fit and the colored curves show the individual
     contributions of the $\pi_1$- and $a_2$-waves.}
 \end{figure}

Reasonably good agreement is achieved between the fit based
on the model described in Sec.~\ref{sec:PWA}  and all data samples.
Exemplarily for all \pbarp\ channels, the result for \PiPiEta\ is shown in Fig.~\ref{fig:pipieta}.
The invariant $\pi^0\eta$ mass as well as the production
and decay angular distribution of the subsystems where the $a_2$- and $\pi_1$-wave are directly contributing are very well described.
In analogy to~\cite{Albrecht:2019ssa} also here a
  non-parametric goodness of fit test has been performed for all three \pbarp\ channels
by utilizing a multivariate analysis based on the concept of statistical energy~\cite{2005NIMPA.537..626A}. The
obtained p-values of 0.405, 0.519 and 0.832 for the channels \PiPiEta, \PiEtaEta\ and \KpKmPi0, respectively,
demonstrate that the quality of the fit is as good as the one without considering the COMPASS data yielding strong
constraints for the $\pi_1$- and $a_2$-wave. In \PiPiEta, the contribution of the $\pi_1$-wave with
(11.9\,$\pm$\,1.6\,$\pm$\,1.9)\%,
of the $a_2$-wave with (30.8\,$\pm$\,2.7\,$\pm$\,1.9)\% and also of
all other waves are in the ballpark of the fractions
obtained by the fit without the COMPASS data (supplemental
  material, Tab. 2). Also the individual contributions in the channels \PiEtaEta\ and
\KpKmPi0\ are similar to the old results.\\
The results for the 11 different \PiPi-scatt\-ering data
sets are
shown in the supplemental material (Fig. 5) and there are no major differences visible
in comparison to~\cite{Albrecht:2019ssa}.\\
\begin{figure*}[htb]
  \centering
  \includegraphics[width=\textwidth]{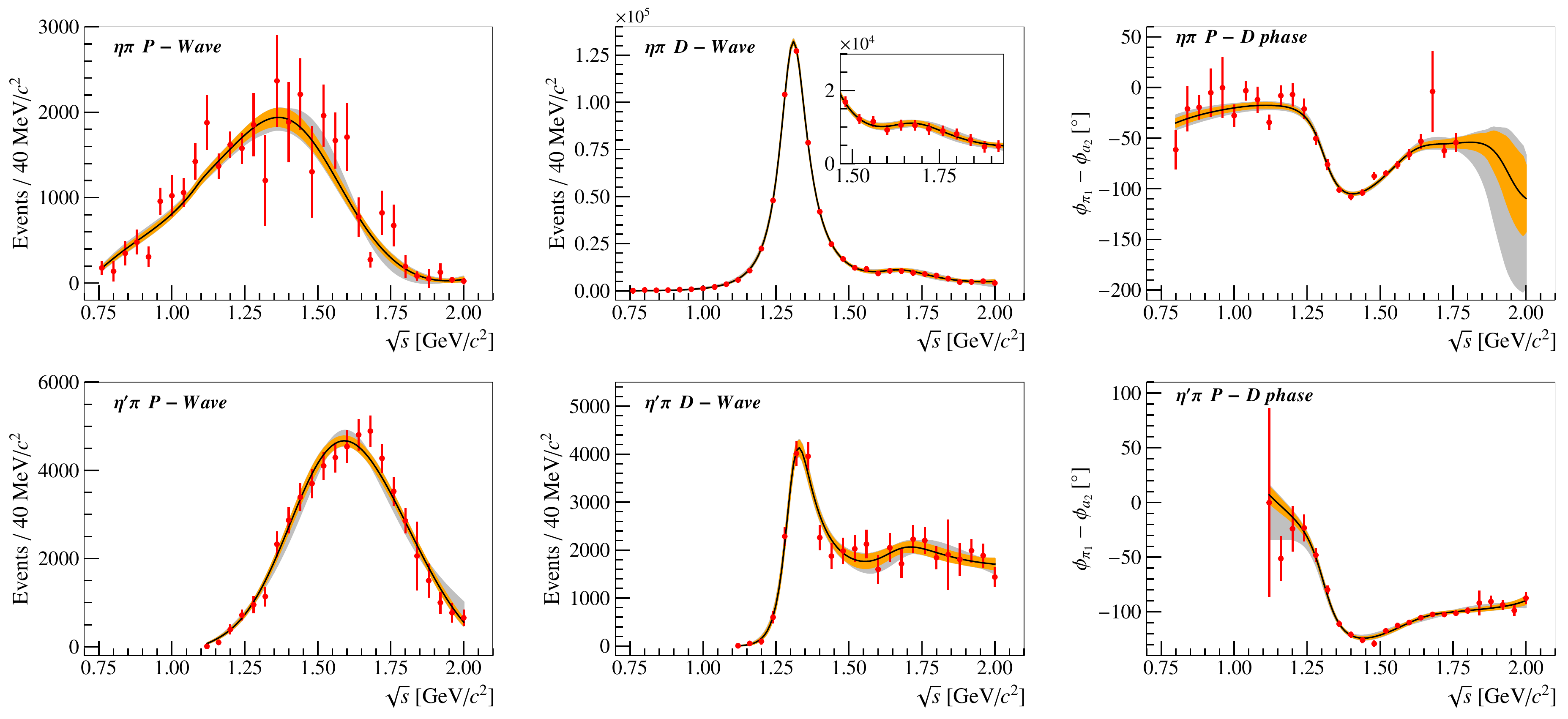}\\
\caption{\label{fig:compass} Fits to the $\pi \eta$
  (upper row) and $\pi \eta^\prime$ (lower row) data from COMPASS.
The intensities of the P- (left), D-wave (center), and their relative
phases (right) are shown. The data are represented by the red
points with error bars. The black curve illustrates our best fit to the data, while
the yellow and gray bands represent the statistical and systematic uncertainty, respectively.}
\end{figure*}
\begin{table*}[htb]%
\caption{\label{tab:poles}%
Obtained masses, total widths and ratios of partial widths for the pole of
the spin-exotic $\pi_1$-wave and for the two poles in
the $a_2$-wave, the $a_2(1320)$ and the $a_2(1700)$. The
  first uncertainty is the statistical and the second the systematic one.
}
\centering
\adjustbox{max width=\linewidth}{
  \begin{tabular}{lrrrr}
    \hline\noalign{\smallskip}
\textrm{name} &
\textrm{pole mass [MeV/$c^2$]} &
\textrm{pole width [MeV]} &
\textrm{$\Gamma_{\pi \eta^\prime}/\Gamma_{\pi \eta}$ [\%]}&
                                                            \textrm{$\Gamma_{KK}/\Gamma_{\pi \eta}$ [\%]}\\
    \hline
    \noalign{\smallskip}
$a_2(1320)$ & 1318.7$\,\pm \,1.9\,^{+1.3}_{-1.3}$ &
                                                   107.5$\,\pm\, 4.6\,^{+3.3}_{-1.8}$
                          &
                            4.6$\,\pm\, 1.5\,^{+7.0}_{-0.6}$ & 31$\,\pm\,22\,^{+9}_{-11}$\\ \noalign{\smallskip}
$a_2(1700)$ & 1686$\,\pm 22\,^{+19}_{-7}$ &
                                            412$\,\pm\,75\,^{+64}_{-57}$
                          & 3.5$\,\pm\,4.4 \,^{+6.9}_{-1.2}$ &
                                                               2.9$\,\pm\, 4.0\,^{+1.1}_{-1.2}$\\\noalign{\smallskip}
$\pi_1$     & \WoUnitsMassPi1  &  \WoUnitsWidthPi1 & 554$\,\pm\,110\,^{+180}_{-27}$ & --\\
  \noalign{\smallskip}
  \hline
\end{tabular}
}
\end{table*}
The comparison between our fit result and the COMPASS data is shown in Fig.~\ref{fig:compass}.
All data are described
remarkably well. It is worth mentioning that
the K-matrix of the $\pi_1$-wave consisting of only one pole can reproduce the shapes of the intensities in
$\pi\eta$ and $\pi\eta^\prime$ even though there is a shift of roughly
200 \mev2c of the peak position
between both channels (Fig.~\ref{fig:compass} (upper left) and (lower
left)). A significantly worse fit result based on the
  information criteria was achieved
for the scenario in which the $\pi_1$-wave in the channel
\pbarpToPi0Pi0Eta\ has been removed from the model. The negative
log-likelihood value increases by more than 125 with
only 20 free parameters less. Similar to the results obtained
in~\cite{Albrecht:2019ssa} also here the $\pi_1$-wave amplitude
is definitely needed for this \pbarp-annihilation channel. Contrary to
the outcome without the $\pi_1$ contribution, the fit taking into account two individual $\pi_1$-poles
does not yield significantly worse results. Based on the chosen
Bayesian and Akaike information criteria~\cite{Albrecht:2019ssa} the
two-pole scenario cannot be completely excluded.\\
The pole positions for the individual resonances described by the K-matrix are extracted in the complex energy plane
of the T-matrix on the Rieman sheet located next to the physical sheet. To some extent also partial widths have been
derived from the residues calculated from the integral
along a closed contour around the pole. The procedure for the
extraction of these properties are explained in detail
in~\cite{Albrecht:2019ssa}. The extracted
resonance parameters for the $\pi_1$  and the two $a_2$ states are
summarized in Tab.~\ref{tab:poles}. The $\pi_1$-mass is significantly
higher compared to the one published
in~\cite{Albrecht:2019ssa}. This is compatible with all
  other findings attributing a lower mass to $\pi_1$, if only
  $\pi\eta$ decays are analyzed. One conjecture is that the
  requirement of unitarity cannot be strictly fulfilled for all
  analyses that take into accout only one decay channel with a weak coupling
    to the resonance.
Apart from a larger width of more than
400 MeV/$c^2$ obtained for the
$a_2(1700)$ resonance, all other masses and widths
are comparable with the ones obtained
  in ~\cite{Rodas:2018owy}. The absolute coupling strengths have not
 been determined because the non-negligible decay channel $\rho \pi$
 is not covered by the fitted data samples. Instead, the ratios
 $\Gamma_{\pi \eta^\prime}/\Gamma_{\pi \eta}$ for all three poles and
$\Gamma_{KK}/\Gamma_{\pi \eta}$ for the $a_2$ resonances have been
determined which should deliver more reasonable results.
The obtained quantities of the remaining resonances can
  be found in Tab. 4 of the supplemental material. The results are in
  the ballpark of other individual measurements~\cite{Zyla:2020zbs}
  and the ones published in~\cite{Albrecht:2019ssa}, except the $f_2$
  state with the highest mass, which
  is located far beyond the phase space of the \pbarp\ and \PiPi\
  data. A general reason for the slight inconsitencies
    compared to the
    outcome of~\cite{Albrecht:2019ssa} is that a large correlation was found
    between the two waves with $\pi_1$ and $a_2$, which are mainly driven by the
    COMPASS data, and the remaining waves representing the \pbarp\ data. The
  results are not meant to supersede the previous ones obtained
  by~\cite{Albrecht:2019ssa}, which use fewer channels and thus less
  parameters. Here a different analysis is presented,
    where in particular the inclusion of the $\pi p$ data leads to much
stronger constraints for the description of the $\pi_1$ and $a_2$
amplitudes. The fit in~\cite{Albrecht:2019ssa} did not include any data for the $\pi\eta^\prime$ decay channel. Also the representation of
the K-matrix for the $a_2$-wave  has been extended by adding the $\pi\eta^\prime$
decay channel.
  
\section{Statistical and Systematic Uncertainties}
The statistical uncertainties are estimated by the boostrap
method~\cite{press2007numerical, EfroTibs93}. Due to the fact that the
coupled-channel fits require a lot of CPU time, only 100
pseudo-data samples are generated and refitted. However, with
this limited number of datasets it is still possible to determine the
standard deviations for each quantity with relative
uncertainties of less than 10\%. The obtained statistical
  uncertainties related to
  the properties of the
  $\pi_1$ and $a_2$ resonances are slightly
  larger than the ones published in~\cite{Rodas:2018owy}, although
  additional \pbarp\ data are used. This is caused by more free parameters needed for the
  description of the two waves. In particular the $a_2$-wave here consists
  of a three channel scenario with the coupling to $\pi\eta$, $\pi\eta^\prime$ and $\bar{K}K$,
  while in~\cite{Rodas:2018owy} the decay to $\bar{K}K$ was not taken
  into accout.\\
The systematic uncertainties listed in Tab.~\ref{tab:poles} are derived from the outcome of alternative fits
which deliver reasonably good results compared to the one with the best
hypothesis by applying the same criteria as
in~\cite{Albrecht:2019ssa}. Also the impact of the K-matrix and P-vector
background terms of the $\pi_1$ pole was investigated and is included in the systematic uncertainties.
In addition the effective transferred momentum squared of
$t_{\text{eff}}= -\,0.1$\,GeV$^2$ for the description of the COMPASS data
has been varied in the range between $-\,0.1$~GeV$^2$ and\linebreak 
$-\,0.5$~GeV$^2$. Only slight differences are obtained which are also considered
as systematic uncertainty.\\
One of the main systematic effects seems to be caused
  by a strong correlation between the widths of the  $\pi_1$ and of the
  $a_2(1700)$ pole which accounts for the relevant large
  uncertainties listed in Tab.~\ref{tab:poles}.

\section{Conclusion}
A coupled-channel analysis of the \pbarp\ annihilation channels \PiPiEta, \PiEtaEta\ and \KpKmPi0\ measured at Crystal Barrel, of
11 different \PiPi-scattering data sets and of the P- and D-waves in the $\pi\eta$ and
$\pi\eta^\prime$ system measured at COMPASS has been performed. The analysis was mainly focused on
the investigation of the spin-exotic $I^G(J^{PC})$ = $1^-(1^{-+})$ wave recently observed in the $\pi\eta$ system of the \pbarp\
channel \PiPiEta\ and in the $\pi\eta$ and $\pi\eta^\prime$
systems of the high-energy $\pi^- p $-scat\-tering data~\cite{Adolph:2014rpp,Adolph:2014rppCorrigendum}.
For a sophisticated description of the dynamics the K-matrix approach
with Chew-Mandelstam functions
has been used which ensures an appropriate consideration of analyticity and
unitarity conditions. The fit can reproduce all 20 different data
samples reasonably well. Only one pole is needed
for an appropriate description of the $\pi_1$-wave amplitude in the two subsystems
$\pi\eta$ and $\pi\eta^\prime$, but a 2-pole scenario cannot be
completely excluded. The mass and width of this single $\pi_1$-pole are measured
to be \MassPi1 and \WidthPi1, respectively. This result is in good agreement with~\cite{Rodas:2018owy} even though a
slightly different description for the dynamics has been chosen and a much larger data base has been exploited. The outcome of the study here confirms
the statement that the two $\pi_1$ resonances listed in the RPP, the 
$\pi_1(1400)$ and $\pi_1(1600)$, might
originate from the same pole. The shift of the pole with
  respect to~\cite{Albrecht:2019ssa} shows that the influence of the $\pi\eta^\prime$
  channel plays an essential role and is in agreement with all previous findings.

\begin{acknowledgements}
  The study was funded by the Collaborative Research Center under the
project CRC 110: {\it Symmetries and the Emergence of Structure in
  QCD}. The authors wish to thank A.~Pilloni and A.~Rodas for the
helpful discussions related to this work. We also gratefully
acknowledge W.~D\"unnweber for providing important insights to the
COMPASS measurement.
Most of the time-consuming fits have been performed on the
Virgo Cluster
at GSI in Darmstadt.
\end{acknowledgements}

\appendix


\begin{thebibliography}{10}
\providecommand{\url}[1]{{#1}}
\providecommand{\urlprefix}{URL }
\expandafter\ifx\csname urlstyle\endcsname\relax
  \providecommand{\doi}[1]{DOI \discretionary{}{}{}#1}\else
  \providecommand{\doi}{DOI \discretionary{}{}{}\begingroup
  \urlstyle{rm}\Url}\fi

\bibitem{Lacock:1996ny}
P.~Lacock, C.~Michael, P.~Boyle, P.~Rowland, Phys. Lett. B \textbf{401}, 308
  (1997).
\newblock \doi{10.1016/S0370-2693(97)00384-5}

\bibitem{Bernard:1997ib}
C.W. Bernard, et~al., Phys. Rev. D \textbf{56}, 7039 (1997).
\newblock \doi{10.1103/PhysRevD.56.7039}

\bibitem{Dudek:2013yja}
J.J. Dudek, R.G. Edwards, P.~Guo, C.E. Thomas, Phys. Rev. D \textbf{88}(9),
  094505 (2013).
\newblock \doi{10.1103/PhysRevD.88.094505}

\bibitem{Woss:2020ayi}
A.J. Woss, J.J. Dudek, R.G. Edwards, C.E. Thomas, D.J. Wilson, Phys. Rev. D
  \textbf{103}(5), 054502 (2021).
\newblock \doi{10.1103/PhysRevD.103.054502}

\bibitem{PhysRevD.65.025012}
A.P. Szczepaniak, E.S. Swanson, Phys. Rev. D \textbf{65}, 025012 (2001).
\newblock \doi{10.1103/PhysRevD.65.025012}.
\newblock \urlprefix\url{https://link.aps.org/doi/10.1103/PhysRevD.65.025012}

\bibitem{Szczepaniak:2006nx}
A.P. Szczepaniak, P.~Krupinski, Phys. Rev. D \textbf{73}, 116002 (2006).
\newblock \doi{10.1103/PhysRevD.73.116002}

\bibitem{Alde:1988bv}
D.~Alde, et~al., Phys. Lett. B \textbf{205}, 397 (1988).
\newblock \doi{10.1016/0370-2693(88)91686-3}

\bibitem{Aoyagi:1993kn}
H.~Aoyagi, et~al., Phys. Lett. B \textbf{314}, 246 (1993).
\newblock \doi{10.1016/0370-2693(93)90456-R}

\bibitem{Thompson:1997bs}
D.R. Thompson, et~al., Phys. Rev. Lett. \textbf{79}, 1630 (1997).
\newblock \doi{10.1103/PhysRevLett.79.1630}

\bibitem{Abele:1998gn}
A.~Abele, et~al., Phys. Lett. B \textbf{423}, 175 (1998).
\newblock \doi{10.1016/S0370-2693(98)00123-3}

\bibitem{Abele:1999tf}
A.~Abele, et~al., Phys. Lett. B \textbf{446}, 349 (1999).
\newblock \doi{10.1016/S0370-2693(98)01544-5}

\bibitem{Salvini:2004gz}
P.~Salvini, et~al., Eur. Phys. J. C \textbf{35}, 21 (2004).
\newblock \doi{10.1140/epjc/s2004-01811-8}

\bibitem{Adams:2006sa}
G.S. Adams, et~al., Phys. Lett. B \textbf{657}, 27 (2007).
\newblock \doi{10.1016/j.physletb.2007.07.068}

\bibitem{Adams:1998ff}
G.S. Adams, et~al., Phys. Rev. Lett. \textbf{81}, 5760 (1998).
\newblock \doi{10.1103/PhysRevLett.81.5760}

\bibitem{Alekseev:2009aa}
M.~Alekseev, et~al., Phys. Rev. Lett. \textbf{104}, 241803 (2010).
\newblock \doi{10.1103/PhysRevLett.104.241803}

\bibitem{Ivanov:2001rv}
E.I. Ivanov, et~al., Phys. Rev. Lett. \textbf{86}, 3977 (2001).
\newblock \doi{10.1103/PhysRevLett.86.3977}

\bibitem{Kuhn:2004en}
J.~Kuhn, et~al., Phys. Lett. B \textbf{595}, 109 (2004).
\newblock \doi{10.1016/j.physletb.2004.05.032}

\bibitem{Lu:2004yn}
M.~Lu, et~al., Phys. Rev. Lett. \textbf{94}, 032002 (2005).
\newblock \doi{10.1103/PhysRevLett.94.032002}

\bibitem{Akhunzyanov:2018lqa}
M.~Aghasyan, et~al., Phys. Rev. D \textbf{98}(9), 092003 (2018).
\newblock \doi{10.1103/PhysRevD.98.092003}

\bibitem{Rodas:2018owy}
A.~Rodas, et~al., Phys. Rev. Lett. \textbf{122}(4), 042002 (2019).
\newblock \doi{10.1103/PhysRevLett.122.042002}

\bibitem{Albrecht:2019ssa}
M.~Albrecht, et~al., Eur. Phys. J. C \textbf{80}(5), 453 (2020).
\newblock \doi{10.1140/epjc/s10052-020-7930-x}

\bibitem{Adolph:2014rpp}
C.~Adolph, et~al., Phys. Lett. B \textbf{740}, 303 (2015).
\newblock \doi{10.1016/j.physletb.2014.11.058}

\bibitem{Adolph:2014rppCorrigendum}
C.~Adolph, et~al., Phys. Lett. B \textbf{811}, 135913 (2020).
\newblock \doi{10.1016/j.physletb.2020.135913}

\bibitem{PhysRevD.91.054008}
D.J. Wilson, J.J. Dudek, R.G. Edwards, C.E. Thomas, Phys. Rev. D \textbf{91},
  054008 (2015).
\newblock \doi{10.1103/PhysRevD.91.054008}.
\newblock \urlprefix\url{https://link.aps.org/doi/10.1103/PhysRevD.91.054008}

\bibitem{Kopf:2014wwa}
B.~Kopf, H.~Koch, J.~Pychy, U.~Wiedner, Hyperfine Interact. \textbf{229}(1-3),
  69 (2014).
\newblock \doi{10.1007/s10751-014-1039-2}

\bibitem{AITCHISON1972417}
I.~Aitchison, Nuclear Physics A \textbf{189}(2), 417 (1972).
\newblock \doi{https://doi.org/10.1016/0375-9474(72)90305-3}.
\newblock
  \urlprefix\url{https://www.sciencedirect.com/science/article/pii/0375947472903053}

\bibitem{PWAinKMatChungBrose}
S.U. Chung, J.~Brose, R.~Hackmann, E.~Klempt, S.~Spanier, C.~Strassburger,
  Annalen der Physik \textbf{507}(5), 404 (1995).
\newblock \doi{https://doi.org/10.1002/andp.19955070504}.
\newblock
  \urlprefix\url{https://onlinelibrary.wiley.com/doi/abs/10.1002/andp.19955070504}

\bibitem{PhysRev.119.467}
G.F. Chew, S.~Mandelstam, Phys. Rev. \textbf{119}, 467 (1960).
\newblock \doi{10.1103/PhysRev.119.467}

\bibitem{Zyla:2020zbs}
P.~Zyla, et~al., PTEP \textbf{2020}(8), 083C01 (2020).
\newblock \doi{10.1093/ptep/ptaa104}

\bibitem{Jackura:2019ccv}
A.W. Jackura, {Studies in Multiparticle Scattering Theory}.
\newblock Ph.D. thesis, Indiana U., Bloomington (main) (2019).
\newblock \doi{10.2172/1570367}

\bibitem{2005NIMPA.537..626A}
B.~{Aslan}, G.~{Zech}, Nuclear Instruments and Methods in Physics Research A
  \textbf{537}, 626 (2005).
\newblock \doi{10.1016/j.nima.2004.08.071}

\bibitem{press2007numerical}
W.~Press, S.~Teukolsky, W.~Vetterling, B.~Flannery, \emph{Numerical Recipes:
  The Art of Scientific Computing}, 3rd edn. (Cambridge University Press,
  2007).
\newblock \urlprefix\url{http://nr.com/}

\bibitem{EfroTibs93}
B.~Efron, R.J. Tibshirani, \emph{An Introduction to the Bootstrap}.
\newblock No.~57 in Monographs on Statistics and Applied Probability (Chapman
  \& Hall/CRC, Boca Raton, Florida, USA, 1993)

\end{thebibliography}
\providecommand{\noopsort}[1]{}\providecommand{\singleletter}[1]{#1}%

\end{document}


\title{Supplemental Material from the Investigation of the Lightest Hybrid Meson Candidate with a
Coupled-Channel Analysis of \pbarp\,-, $\pi^- p$\,- and \PiPi-Data}

\maketitle

\section{Results}

All obtained results are summarized by the following tables and
figures. The relevant details related to the extraction of the
resonance
properties, cross sections and the histograms can be found in the publication of the recent coupled-channel analysis of \pbarp\,-annihilation and \PiPi-scattering data~\cite{Albrecht:2019ssa}.

\begin{table}[htb]
\caption{Likelihood values and the number of free 
  parameters for fits with the best and alternative hypotheses.
  $\Delta$NLL, $\Delta$ndf, $\Delta$BIC and $\Delta$AIC are the differences of the
obtained negative log-likelihood values (NLL), the number of
free parameters (NFP)
as well as the BIC and AIC values~\cite{Anderson:2004} between the alternative and the best
hypothesis. The fits marked with (*) are taken into account for the
estimation of the systematic uncertainties. The $\phi(1680)$ and also
the $\rho_3(1690)$, which are each taken into account in one alternative
fit, are described by  a Breit-Wigner function.}

\label{tab:FitWCOMPASSLHs}      
\footnotesize
\centering
\adjustbox{max width=\linewidth}{
  \begin{tabular}{l  r r r r r r}
    \\
  \hline\hline\noalign{\smallskip}
 hypothesis   & NLL & NFP & $\Delta$NLL & $\Delta$NFP & $\Delta$BIC & $\Delta$AIC\\
\noalign{\smallskip}
  \hline \noalign{\smallskip}
  best hypothesis & $-$43156 & 617 & 0 & 0 & 0  & 0 
  \\ \noalign{\smallskip}
  w/  3$\,\rho$ poles (*) & 	$-$43211 & 635 & $-$55 & 18 & 99 & $-$76
    \\ \noalign{\smallskip}
	w/  2$\,\pi_1$ poles (*)& $-$43189 & 636 & $-$33 & 19 & 155 & $-$29                                      
    \\  \noalign{\smallskip}
  w/  $\phi(1680) \; \pi^0$ (*) & $-$43192 & 641 & $-$36 & 24 & 208 & $-$31
 \\  \noalign{\smallskip}
    w/  $\pi_1 \; \eta$ (*) & 	$-$43181 & 638 & $-$25 & 21 & 195 & $-$2
  \\  \noalign{\smallskip}
   w/ $\rho_3(1690) \; \pi^0$ (*) & $-$43203 & 655 & $-$47 & 38 & 350 & $-$25
  \\ 
    \\ 
  w/ 1 $a_2$ pole & 	$-$43032 & 597 & 124 & $-$20 & 14 & 214
  \\ \noalign{\smallskip}
  w/o  $\pi_1 \; \pi^0$  & $-$43031 & 597 & 125 & $-$20 & 15 & 203
  \\  \noalign{\smallskip} 
     w/ 3 $f_2$ poles  & 	$-$42812 & 583 & 344 & $-$34 & 173 & 600
 \\  \noalign{\smallskip} 
   w/ 1 $a_0$ pole & $-$42873 & 594 & 283 & $-$23 & 296 & 519
   \\  \noalign{\smallskip} 
  w/ 4 $f_0$ poles & $-$42713 & 591 & 443 & $-$26 & 581 & 833
  \\ \noalign{\smallskip}
  \hline\hline
\end{tabular}
}
\end{table}

\begin{table}[htb]
  \caption{Contributions in $\%$ of the individual waves for the three
    channels \pbarpToPi0Pi0Eta
	, \PiEtaEta\
    and \KpKmPi0.
  }

\label{tab:FitFractions}      
\setlength{\tabcolsep}{3pt}
\footnotesize
\centering
\adjustbox{max width=\linewidth}{
  \begin{tabular}{l r r r}
    \\
  \hline\hline \noalign{\smallskip}
         & \multicolumn{3}{c}{contribution (in $\%$) for channel} \\
  component       & \PiPiEta\ & \PiEtaEta\ & \KpKmPi0\ \\
  \noalign{\smallskip}
  \hline
  \noalign{\smallskip} 
$f_0\pi^0$                 & --                         & 33.7~$\pm$~3.7$\pm$~5.2   & 14.1~$\pm$~2.1$\pm$~2.3  \\
$f_0\eta $                 & 14.6~$\pm$~2.2$\pm$~0.9   & --                         & --                        \\
   \noalign{\smallskip}                                                                                           
$f_2\pi^0$                 & --                         &
                                                          23.9~$\pm$~4.0$\pm$~3.1   & 21.2~$\pm$~3.0$\pm$~4.3  \\
$f_2\eta $                 & 50.0~$\pm$~3.9$\pm$~2.8   & --                         & --                        \\
   \noalign{\smallskip}                                                                                           
$\rho\pi^0$                & --                         & --                         & 13.2~$\pm$~2.6$\pm$~2.1  \\
   \noalign{\smallskip}                                                                                           
$a_0\pi^0$                 & 20.3~$\pm$~2.6$\pm$~2.3   & --                         & 9.5~$\pm$~1.7$\pm$~1.5   \\
$a_0\eta $                 & --                         & 36.0~$\pm$~6.0$\pm$~4.8   & --                        \\
   \noalign{\smallskip}                                                                                           
$a_2\pi^0$                 & 30.8~$\pm$~2.7$\pm$~1.9   & --                         & 2.2~$\pm$~0.7$\pm$~0.9   \\
$a_2\eta $                 & --                         & 27.6~$\pm$~3.8$\pm$~4.2   & --                        \\
   \noalign{\smallskip}                                                                                           
$K^*(892)^{\pm}K^{\mp}  $  & --                         & --                         & 37.6~$\pm$~2.8$\pm$~1.8  \\
$(K\pi)^{\pm}_{S}K^{\mp}$  & --                         & --                         & 9.0~$\pm$~2.2$\pm$~2.8   \\
$\phi(1020)\pi^{0}      $  & --                         & --                         & 2.8~$\pm$~0.5$\pm$~0.2   \\
$\pi_{1}\pi^0           $  & 11.9~$\pm$~ 1.6$\pm$~1.9   & --                         & --                        \\
   \noalign{\smallskip}
  \hline
  \noalign{\smallskip}  
$\Sigma$                   & 127.9~$\pm$~7.7$\pm$~5.3  & 121~$\pm$~11$\pm$~4  & 109.7~$\pm$~4.5$\pm$~3.7 \\
  \hline \hline
\end{tabular}
}
\end{table}

\begin{figure}[b]
\hspace{3mm}  \includegraphics[width=\textwidth]{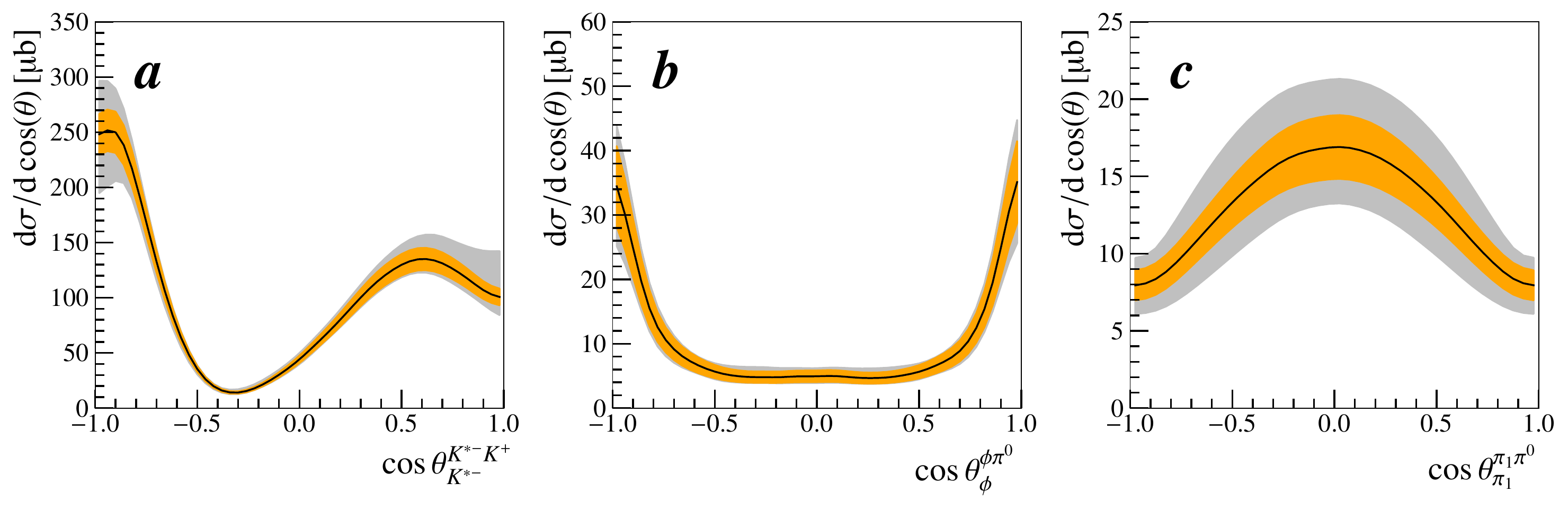}\\[-4ex]
\caption{The differential production cross sections 
for the resonances $K^*(892)^-$ (a), $\phi(1020)$ (b) in the
reaction \pbarpToKpKmPi0\ and for
the spin-exotic $\pi_1$ wave (c) in the reaction
\pbarpToPi0Pi0Eta\ derived from the final fit
  result. The amplitudes for the two
possible  $\pi^0 \eta$ subsystems of the $\pi_1$ decay are summed
coherently. The black curves represent the fit result, the yellow bands
illustrate the statistical uncertainty and the gray 
bands stand for the systematic uncertainty obtained from the alternative fits.}
\label{fig:crossX}
\end{figure}

\begin{table}
\caption{Production cross sections for the resonances $K^*(892)^\pm$, $\phi(1020)$ in the
reaction \pbarpToKpKmPi0\ and for the spin-exotic $\pi_1$ wave in the reaction \pbarpToPi0Pi0Eta\ . 
The first error is the statistical and the second one the systematic uncertainty from this
analysis. The third error represents the uncertainty for the total cross section of the reaction 
$\bar{p} p \to K^{+}K^{-}\pi$ and $\bar{p} p \to
\pi^{0}\pi^{0}\eta$. The cross section of the $\pi_1$ wave
  is
  calculated by the coherent sum of the amplitudes for the two
possible  $\pi^0 \eta$ subsystems. }
\footnotesize
\centering
\begin{tabular}{lr}
  \\
\hline \hline
   \noalign{\smallskip}         
process                   & cross section [$\upmu$b] \\
   \noalign{\smallskip}         
\hline
   \noalign{\smallskip}         
$\bar{p}p \to K^*(892)^{\pm}K^{\mp}                   $  &  396~$\pm$~29$\pm$~19$\pm$~42 \\
   \noalign{\smallskip}         
$\bar{p}p \to \phi(1020)\pi^{0}                       $  & 20.2~$\pm$~3.6$\pm$~1.4$\pm$~2.1     \\
   \noalign{\smallskip}         
$\bar{p}p \to \pi_{1}\pi^0,\pi_{1}\to\pi^0\eta        $  & 25.8~$\pm$~3.5$\pm$~4.1$\pm$~1.5     \\
   \noalign{\smallskip}         
\hline \hline
\end{tabular}
\end{table}

\begin{figure}[ht]
\includegraphics[width=0.5\textwidth]{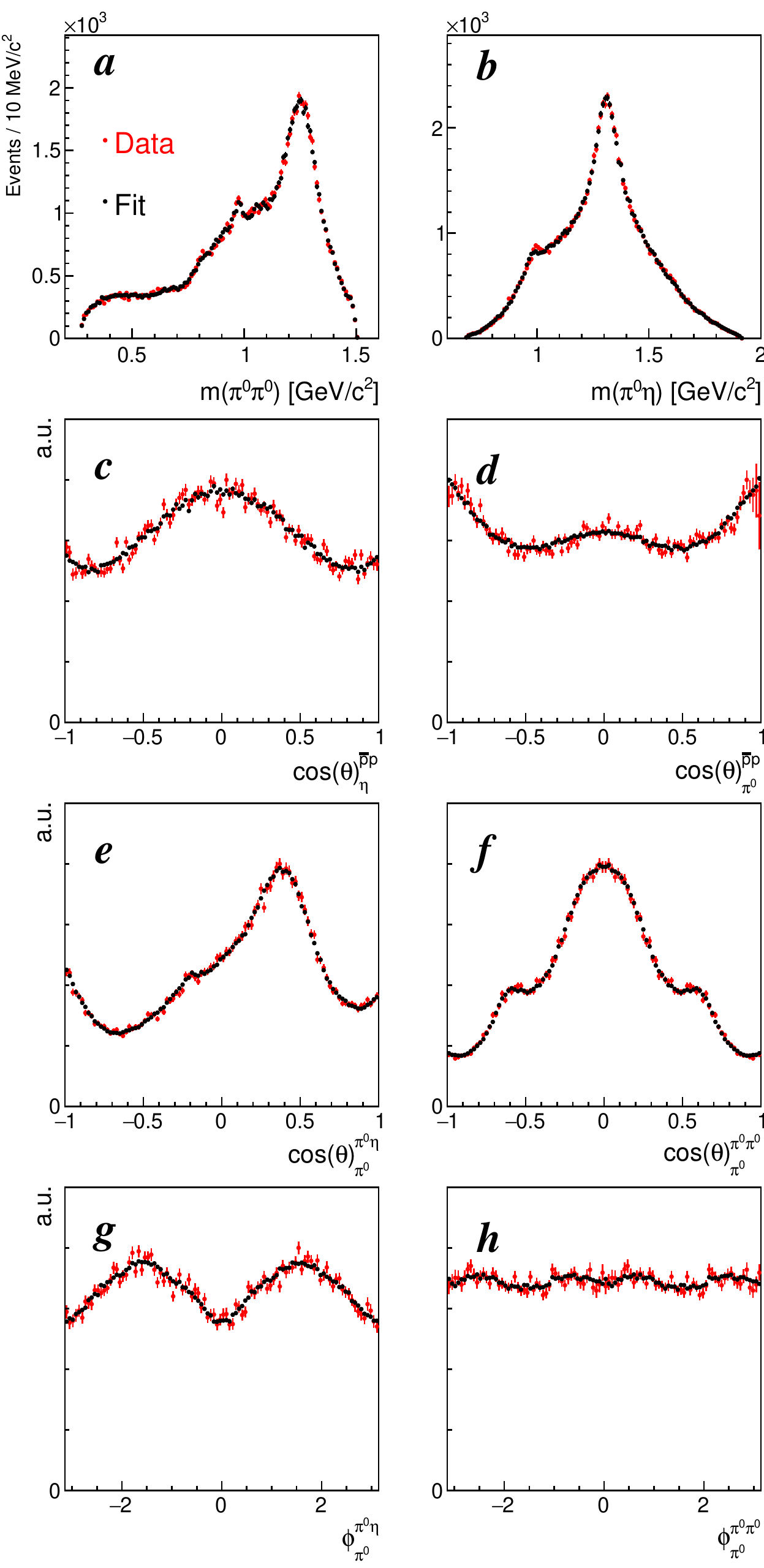}
\centering
\caption{Not efficiency-corrected invariant $\pi^0\pi^0$ and $\pi^0\eta$ mass distribution
  (a and b), efficiency-corrected decay angular distributions for the
  production (c and d) and for the decay (e to h) of the reaction
  \pbarpToPi0Pi0Eta. The data are represented by red points with error bars and the
 fit result is illustrated by the black points.}
\label{fig:ResultPiPiEta}       
\end{figure}

\begin{figure}[ht]
  \includegraphics[width=0.5\textwidth]{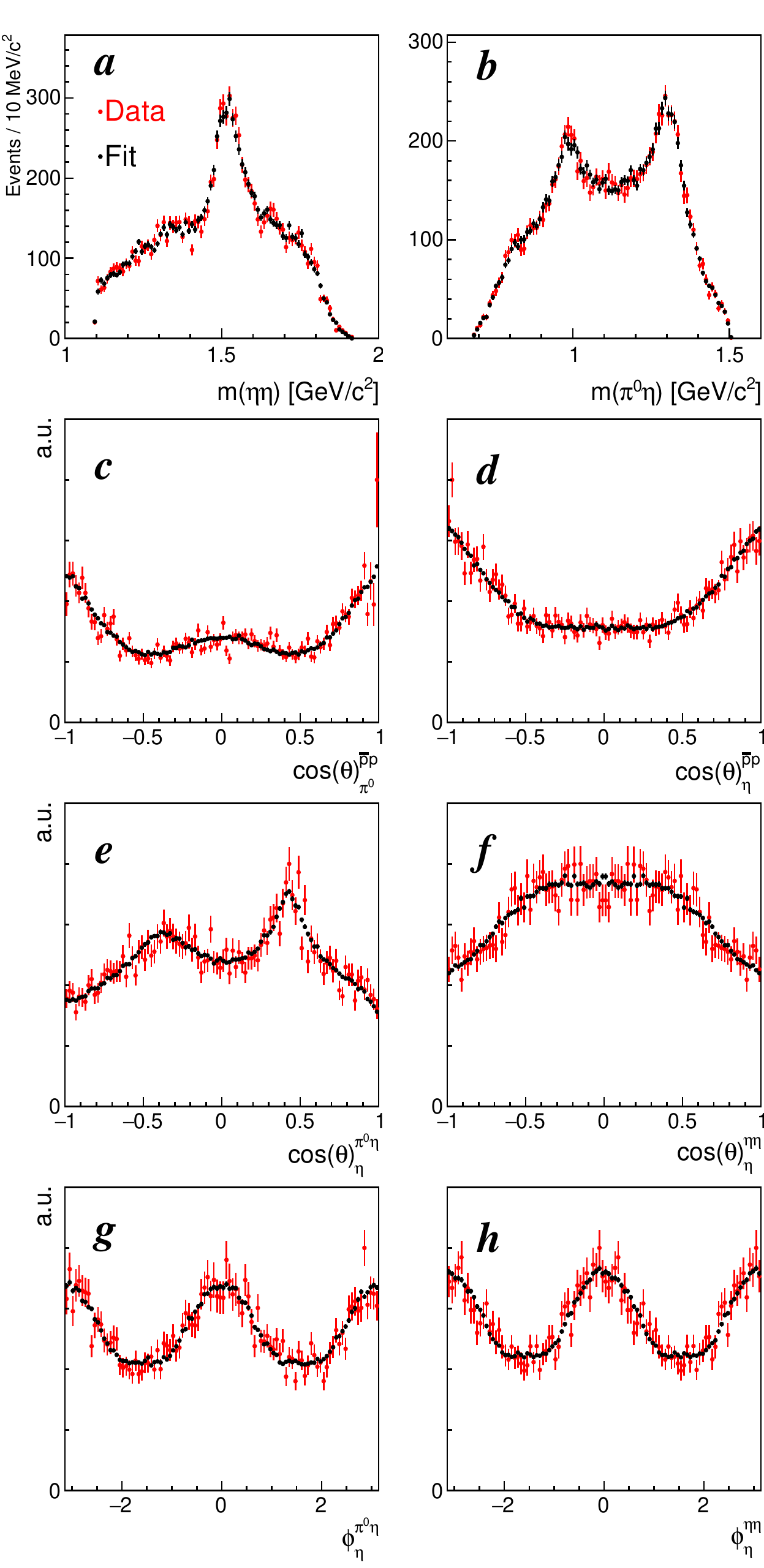}
\centering
\caption{Not efficiency-corrected invariant $\eta\eta$ and $\pi^0\eta$ mass distribution
  (a and b), efficiency-corrected decay angular distributions for the
  production (c and d) and for the decay (e to h) of the reaction
  \pbarpToPiEtaEta. The data are represented by red points with error bars  and the
 fit result is illustrated by the black points.}
\label{fig:ResultPiEtaEta}       
\end{figure}

 \begin{figure}[ht]
  \includegraphics[width=0.5\textwidth]{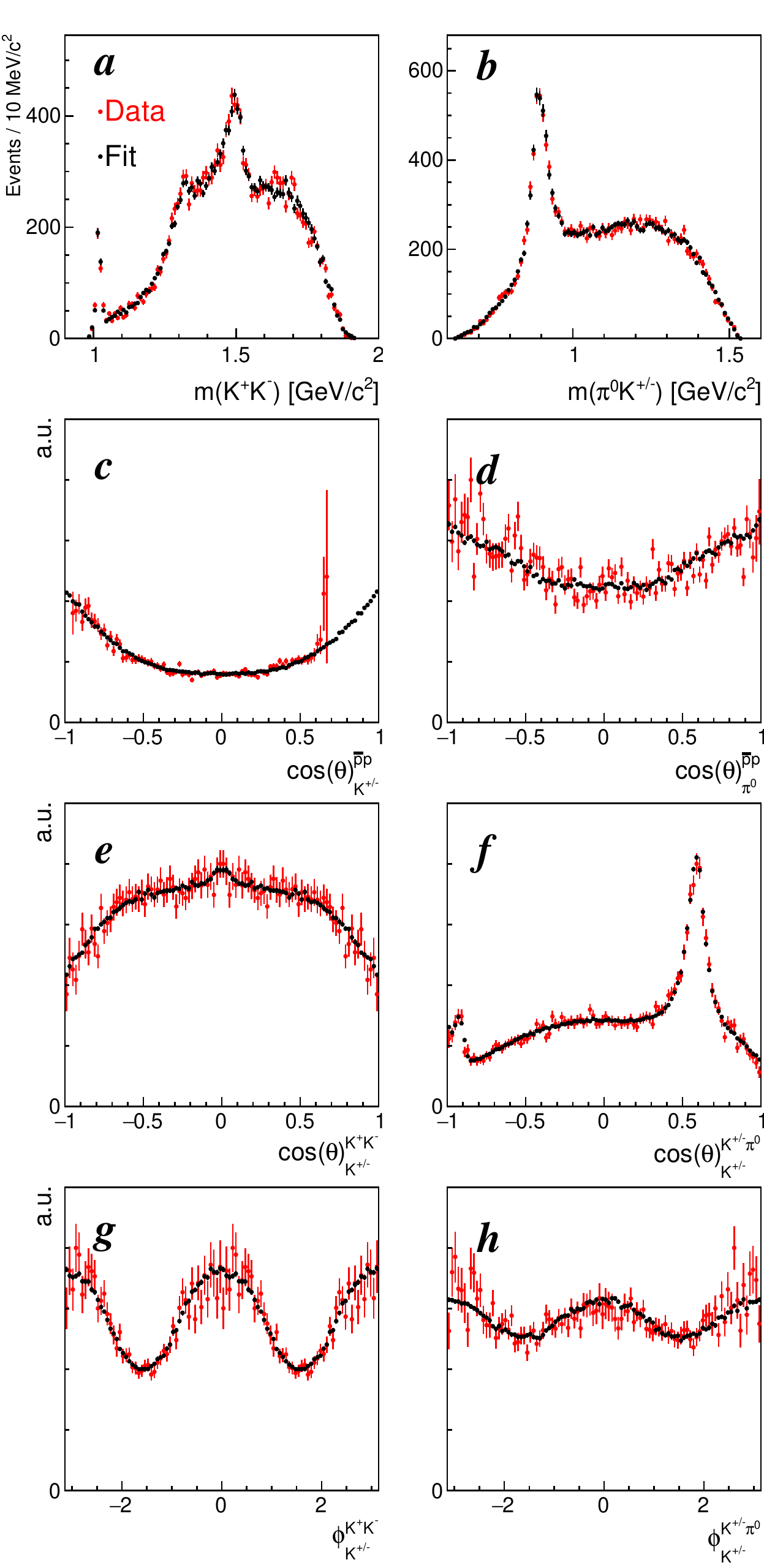}
\centering
\caption{Not efficiency-corrected invariant $K^+K^-$ and $K^\pm\pi^0$ mass distribution
  (a and b), efficiency-corrected decay angular distributions for the
  production (c and d) and for the decay (e to h) of the reaction
  \pbarpToKpKmPi0. The data are represented by red points with error bars  and the
 fit result is illustrated by the black points.}
\label{fig:ResultKKPi}       
\end{figure}

\begin{table*}[htb]
\caption{ Masses, widths and branching fractions for the individual
  resonances. For the isolated resonances $K^*(892)^\pm$ and
    $\phi(1680)$, the values of the Breit-Wigner parameters are
  listed. The $\phi(1020)$ has been described by a
  non-relativistic Voigtian.
  For the $\pi_1$ and all $f_0$, $f_2$, $a_0$, $a_2$ and $\rho$-states the pole
  positions are extracted from the T-matrix. Absolute and relative branching fractions are 
  extracted for the $f_2(1270)$, $f_{2}^{\prime}(1525)$,
  $\rho(1700)$ and for the
  $a_0$, $a_2$ and $\pi_1$ resonances, respectively. 
 The quantities for the $a_0(980)$ and $f_0(980)$ mesons are
  determined on the two 
  relevant Riemann sheets below and above the $K\bar{K}$ threshold
  individually. The sheets of the complex plane are defined by
  the signs ($+$ and $-$) of the imaginary part of the channel
  momenta. The order of the channels for the $f_0(980)$ resonance is
  $\pi \; \pi$, $2\pi \, 2\pi$, $K\;\bar{K}$, $\eta \; \eta$ and $\eta
\; \eta^\prime$ and for the $a_0(980)$ resonance it is $\pi \; \eta$ and $K\;\bar{K}$. 
  The statistical uncertainty estimated with the
    bootstrap method is given by the first and the 
  systematic uncertainty is provided by the second error.
    Each  resonance is assigned to the data samples which
   are contributing significantly to the extraction of the resonance
   properties: \pbarp\ data
  (\pbarp), COMPASS ($\pi p$)  or scattering data (scat).
    The obtained properties of the 
  $f_0(500)$ and $\rho(770)$ states  are not explicitly listed, because they are almost exclusively driven by the
  $\pi\pi$-scattering data from the model-independent calculations~\cite{GarciaMartin:2011cn}. 
}

\label{tab:Properties}      
\footnotesize
\centering
\begin{tabular}{l l r r r r r }
  \\ 
  \cline{1-4} \noalign{\vskip 0.6mm} \cline{1-4} \noalign{\smallskip}
 name           & relevant data & Breit-Wigner mass   & Breit-Wigner width & & &\\ 
                &         &   [\mev2c] &   $\Gamma$ [MeV] &  & & \\
\cline{1-4}\noalign{\smallskip} 
$K^*(892)^\pm$       & \pbarp & 893.8~$\pm$~1.0$\pm$~0.8
                                                      &
                                                        56.3~$\pm$~2.0$\pm$~1.0    & & & \\
  \noalign{\smallskip}
$\phi(1020)$    & \pbarp & 1018.4~$\pm$~0.5$\pm$~0.2 & 4.2~(fixed)      & & & \\
  \noalign{\smallskip} \noalign{\smallskip}
  \cline{1-4}
  \noalign{\smallskip}
  name     & relevant data & pole mass  & pole width  &  & & \\
           &             &   [\mev2c] & $\Gamma$ [MeV] &  & & \\
  \noalign{\smallskip}
  \cline{1-4}\noalign{\smallskip}
$f_{0}(980)^{---++}$    &  scat  &
                                   977.8~$\pm$~0.6$\pm$~1.4 & 98.8~$\pm$~6.6$\pm$~11.2    & & & \\
  \noalign{\smallskip}
$f_{0}(980)^{--+++}$    &  scat  &
                                   992.6~$\pm$~0.3$\pm$~0.5 & 61.2~$\pm$~1.2$\pm$~1.7    & & & \\
  \noalign{\smallskip}
$f_{0}(1370)$  &   scat  & 1281~$\pm$~11~$\pm$~26 &
                                                                 410~$\pm$~12$\pm$~50  & & & \\
  \noalign{\smallskip}
$f_{0}(1500)$  & \pbarp\ + scat  &
                                   1511.0~$\pm$~8.5~$^{+3.5}_{-14.0}$ & 81.1~$\pm$~4.5~$^{+26.9}_{-0.5}$    & & & \\
  \noalign{\smallskip}
$f_{0}(1710)$   & \pbarp\ + scat  & 1794.3~$\pm$~6.1~$^{+47.0}_{-61.2}$ & 281~$\pm$~32~$^{+12}_{-80}$    & & & \\
  \noalign{\smallskip}
  \noalign{\smallskip}
$f_{2}(1810)$   & scat   & 1769~$\pm$~26~$^{+3}_{-26}$ &
                                                                      201~$\pm$~57~$^{+13}_{-87}$ & & & \\
  \noalign{\smallskip}
$f_{2}(X)$   & scat   & 2119.9~$\pm$~6.4~$^{+25.7}_{-1.1}$ & 343~$\pm$~11~$^{+32}_{-11}$ & & & \\
  \noalign{\smallskip} \noalign{\smallskip}
  \cline{1-5}
  \noalign{\smallskip}
  name   & relevant data & pole mass  & pole width  & $\Gamma_{\pi \eta^\prime}/\Gamma_{\pi\eta}$  & & \\
                &                   &   [\mev2c] & $\Gamma$ [MeV] & $[\%]$ & & \\
  \noalign{\smallskip}
  \cline{1-5}
  \noalign{\smallskip}
$\pi_{1}$  & \pbarp\ + $\pi p$ & 1623~$\pm$~47~$^{+24}_{-75}$ & 455~$\pm$~88~$^{+144}_{-175}$    & 554~$\pm$~110~$^{+180}_{-27}$   & & \\
 \noalign{\smallskip} \noalign{\smallskip}
  \cline{1-5}
  \noalign{\smallskip}
  name   & relevant data & pole mass  & pole width  & $\Gamma_{KK}/\Gamma_{\pi \eta}$  & & \\
             &                   &   [\mev2c] & $\Gamma$ [MeV] & $[\%]$ & & \\
  \noalign{\smallskip}
    \cline{1-5}\noalign{\smallskip}
$a_{0}(980)^{--}$  & \pbarp & 1002.7~$\pm$~8.8$\pm$~4.2 &
                                                                       132~$\pm$~11$\pm$~8    & 14.8~$\pm$~7.1$\pm$~3.6   & & \\
  \noalign{\smallskip}
$a_{0}(980)^{-+}$  & \pbarp & 1003.3~$\pm$~8.0$\pm$~3.7 &
                                                                       101.1~$\pm$~7.2$\pm$~3.0    & 13.5~$\pm$~6.2$\pm$~3.1   & & \\
  \noalign{\smallskip}
$a_{0}(1450)$  & \pbarp & 1303.0~$\pm$~3.8$\pm$~1.9 & 109.0~$\pm$~5.0$\pm$~2.9   & 396~$\pm$~72$\pm$~72    & & \\
  \noalign{\smallskip} \noalign{\smallskip}
  \cline{1-6}
  \noalign{\smallskip}
  name   & relevant data & pole mass  & pole width  & $\Gamma_{KK}/\Gamma_{\pi\eta}$ & $\Gamma_{\pi\eta^\prime}/\Gamma_{\pi\eta}$ &  \\
             &                   &   [\mev2c] & $\Gamma$ [MeV] & $[\%]$ & $[\%]$ & \\
 \noalign{\smallskip}
 \cline{1-6}\noalign{\smallskip}
$a_{2}(1320)$  & \pbarp\ + $\pi p$ &
                                     1318.7~$\pm$~1.9~$^{+1.3}_{-1.3}$ & 107.5~$\pm$~4.6~$^{+3.3}_{-1.8}$    & 31~$\pm$~22~$^{+9}_{-11}$ & 4.6~$\pm$~1.5~$^{+7.0}_{-0.6}$ & \\
  \noalign{\smallskip}
$a_{2}(1700)$   & \pbarp\ + $\pi p$ & 1686 ~$\pm$~22~$^{+19}_{-7}$ & 412~$\pm$~75~$^{+64}_{-57}$ & 2.9~$\pm$~4.0~$^{+1.1}_{-1.2}$   & 3.5~$\pm$~4.4~$^{+6.9}_{-1.2}$ &  \\
 \noalign{\smallskip} \noalign{\smallskip}
  \hline
 \noalign{\smallskip} 
  name & relevant data  & pole mass & pole width  & $\Gamma_{\pi\pi}/\Gamma $ & $\Gamma_{KK}/\Gamma$ & $\Gamma_{\eta\eta}/\Gamma $ \\
           &            &   [\mev2c]        & $\Gamma$ [MeV]  & $ [\%]$ & $[\%]$ & $ [\%]$ \\ 
  \noalign{\smallskip}\hline\noalign{\smallskip}
  $f_{2}(1270)$  & \pbarp\ + scat & 1262.4~$\pm$~0.2~$^{+0.2}_{-0.3}$
                                                      &
                                                        168.0~$\pm$~0.7~$^{+1.7}_{-0.1}$  & 87.7~$\pm$~0.3~$^{+4.8}_{-4.4}$  & 2.6~$\pm$~0.1~$^{+0.1}_{-0.2}$   & 0.3~$\pm$~0.1~$^{+0.0}_{-0.1}$ \\
  \noalign{\smallskip}
$f_{2}^{\prime}(1525)$ & \pbarp\ + scat  &
                                           1514.7~$\pm$~5.2~$^{+0.3}_{-7.4}$  & 82.3~$\pm$~5.2~$^{+11.6}_{-4.5}$   & 2.1~$\pm$~0.3~$^{+0.8}_{-0.0}$  & 67.2~$\pm$~4.2~$^{+5.0}_{-3.8}$    & 9.8~$\pm$~3.8~$^{+1.7}_{-3.3}$ \\
  \noalign{\smallskip}
\noalign{\smallskip}
  \noalign{\smallskip}
   $\rho(1700)$ & \pbarp\ + scat & 1700~$\pm$~27~$^{+13}_{-16}$ & 181~$\pm$~25~$^{+0.0}_{-16}$   & 13.6~$\pm$~1.2~$^{+0.9}_{-0.5}$  & 0.8~$\pm$~0.1~$^{+0.0}_{-0.0}$ & -  \\
\noalign{\smallskip}
  \hline \hline
\end{tabular}
\end{table*}

\clearpage
\begin{figure*}[b]
\includegraphics[width=\textwidth]{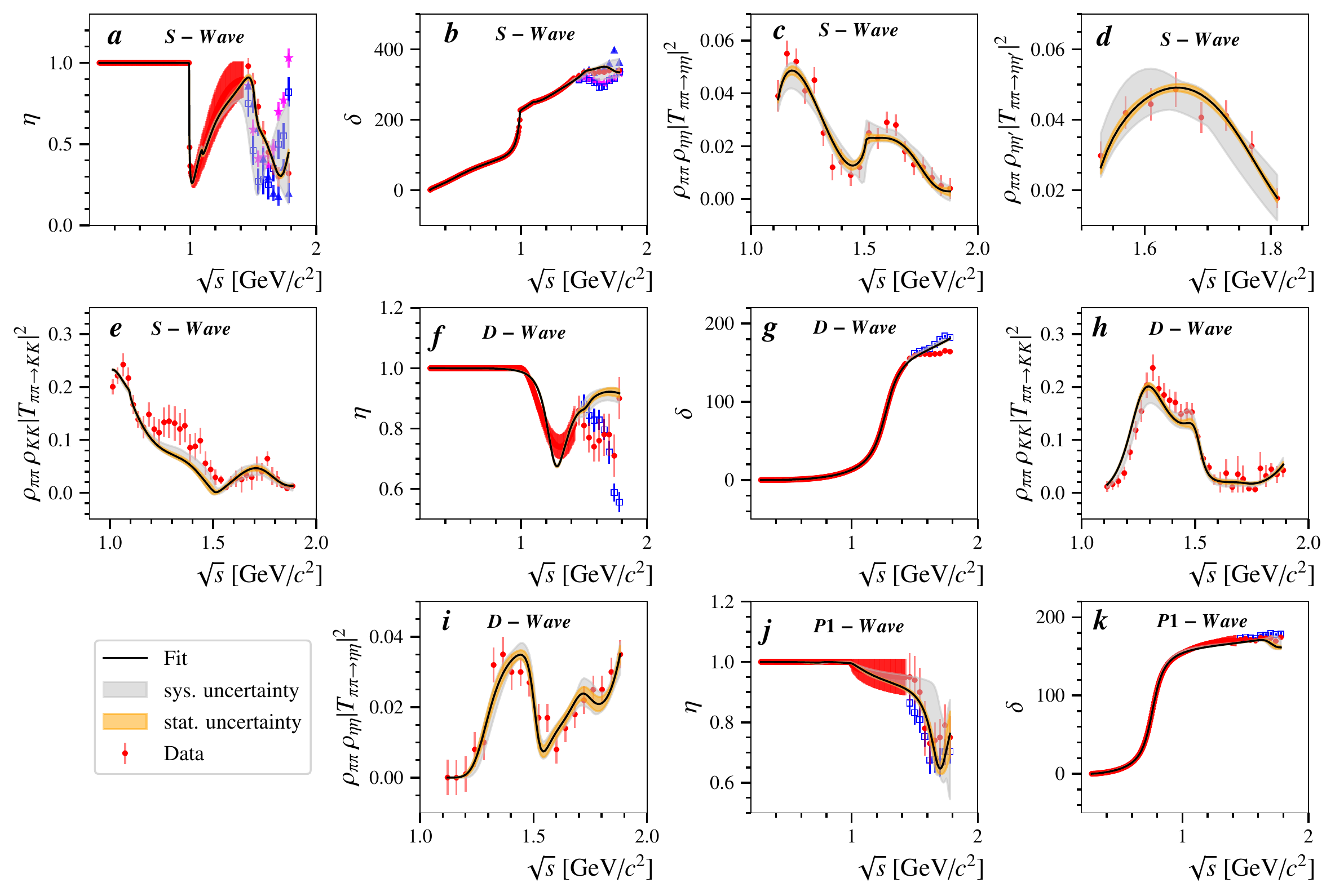}\\[-4ex]
\caption{Results for scattering data. Figs.\,(a and b) show the
  inelasticity and phase shift of the reaction $\pi\pi\to\pi\pi$, 
(c to e) the $|T|^2$-values of the processes
$\pi\pi\to\eta\eta,\eta\eta',KK$ for the S-wave. Figs.\,(f and g) display the inelasticity and phase shift of $\pi\pi\to\pi\pi$ for the D-wave and
(h and i) the $|T|^2$-values of $\pi\pi\to KK,
\eta\eta$ for the D-wave. Fig.\,(j) shows the inelasticity and (k) the phase shift of $\pi\pi\to\pi\pi$
for the P-wave. The data considered in the best fit are given by
red points with error bars which includes the solution (-~-~-) from
~\cite{Ochs:2013gi} (a and b), and solution (-~-~-) from ~\cite{Hyams:1975mc} (f, g, j, k). 
Data sets from multiple solutions are taken into account as alternative fits. 
Data points for solution (-~+~-) from ~\cite{Ochs:2013gi} are labeled
with magenta stars (a and b), and data points for the
solution (-~-~-) and solution (-~+~-)
from ~\cite{Hyams:1975mc}  are labeled with blue triangles
(a and b) and blue squares (a, b, f, g, j, k). The black curves represent the fit result, the tiny yellow bands 
   illustrate the statistical uncertainty and the gray
   bands stand for the systematic uncertainty obtained from the
   alternative fits. 
}
\label{fig:ResultsScattering}      
\end{figure*}

\clearpage
\section{K-matrix parameters}
The following tables list the K-matrix parameters for the waves as
obtained in our best fit to the data. All background parameters are
fixed to 0 for those channels where no data are available and which
are only taken into account as effective channels for the fulfillment
of the unitarity. All units here are given in GeV. 

\begin{table}[h]
  \caption{K-matrix parameters for the $f_{0}$-wave. Only constant
    background terms are taken into account. Indices 0 to
    4 stand for
    the channels
    $\pi\pi,2\pi\,2\pi,K\bar{K},\eta\eta$ and $\eta\eta^{\prime}$. For the
    Chew-Mandelstam function, the $2\pi\,2\pi$ channel is treated as
    an effective $2\pi\,2\pi$ channel as described in~\cite{Albrecht:2019ssa}. The
    $\pi$-mass has been set to $m_{\pi^0}$, the $K$-mass to
    $m_{K^\pm}$ and the $\bar{K}$-mass to $m_{\bar{K}^0}$.}
  \label{tab:kmatrixf0}
  \centering
	\begin {tabular}{ccccccc} \\
	\hline\hline
	pole name $\alpha$&  $m^{\text{bare}}_{\alpha}$ & $g^{\text{bare}}_{\alpha \; \pi\pi}$ & $g^{\text{bare}}_{\alpha \;  2\pi\,2\pi}$ & $g^{\text{bare}}_{\alpha \; K\bar{K}}$ & $g^{\text{bare}}_{\alpha \; \eta\eta}$ & $g^{\text{bare}}_{\alpha \; \eta\eta^{\prime}}$  \\
          \hline

 $f_{0}(500)           $  & 0.51461     &  0.74987  & $-$0.01257 & 0.27536  & $-$0.15102 &  0.36103  \\  
 $f_{0}(980)           $  & 0.90630     &  0.06401  &  0.00204 & 0.77413  &  0.50999 &  0.13112  \\  
 $f_{0}(1370)          $  & 1.23089     & $-$0.23417  & $-$0.01032 & 0.72283  &  0.11934 &  0.36792  \\  
 $f_{0}(1500)          $  & 1.46104     &  0.01270  &  0.26700 & 0.09214  &  0.02742 & $-$0.04025  \\  
 $f_{0}(1710)          $  & 1.69611     & $-$0.14242  &  0.22780 & 0.15981  &  0.16272 & $-$0.17397  \\  
 \hline
	\end {tabular}
	\begin {tabular}{cccccccccc} 
    \noalign{\smallskip}
          \multicolumn{2}{c}{background terms} \\
	$\tilde{c}_{00}$ &  0.03728   & & & & & & & &\\
	$\tilde{c}_{01}$ &  0   & $\tilde{c}_{11}$ & 0 & & & & & & \\
	$\tilde{c}_{02}$ & $-$0.01398   & $\tilde{c}_{12}$ & 0 & $\tilde{c}_{22}$ &  0.02349  & & & & \\
	$\tilde{c}_{03}$ & $-$0.02203   & $\tilde{c}_{13}$ & 0 & $\tilde{c}_{23}$ &  0.03101  & $\tilde{c}_{33}$ & $-$0.13769 &  & \\
	$\tilde{c}_{04}$ &  0.01397   & $\tilde{c}_{14}$ & 0 & $\tilde{c}_{24}$ & $-$0.04003  & $\tilde{c}_{34}$ & $-$0.06722 & $\tilde{c}_{44}$ & $-$0.28401 \\
	\hline
	\end {tabular}
	
	\begin {tabular}{cc}
    \noalign{\smallskip}
  \multicolumn{2}{c}{Adler zero terms} \\
	$s_0$ & 0.0091125  \\
	$s_{\text{norm}}$ & 1  \\
	\hline\hline
	\end {tabular}
      \end{table} 
  \begin{table}[h!]
    \caption{K-matrix parameters for the $f_{2}$-wave. Only constant
      background terms are taken into account. Indices 0
      to 3 stand for the
      channels $\pi\pi,2\pi\,2\pi,K\bar{K}$ and $\eta\eta$. For the
      Chew-Mandelstam function, the $2\pi\,2\pi$ channel is treated as an
      effective $2\pi\,2\pi$ channel as described in~\cite{Albrecht:2019ssa}. The
      $\pi$-mass has been set to $m_{\pi^0}$, the $K$-mass to
      $m_{K^\pm}$ and the $\bar{K}$-mass to $m_{\bar{K}^0}$.}
    \label{tab:kmatrixf2}
  \centering
	\begin {tabular}{cccccc} \\
	\hline\hline
	pole name $\alpha$&  $m^{\text{bare}}_{\alpha}$ & $g^{\text{bare}}_{\alpha \; \pi\pi}$ & $g^{\text{bare}}_{\alpha \; 2\pi\,2\pi}$ & $g^{\text{bare}}_{\alpha \; K\bar{K}}$ & $g^{\text{bare}}_{\alpha \; \eta\eta}$  \\
          \hline
 $f_{2}(1270)          $  & 1.15299     &  0.40033  & 0.15479   & $-$0.08900   &   $-$0.00113    \\              
 $f_{2}^{\prime}(1525) $  & 1.48359     &  0.01820  & 0.17300   &  0.32393   &   0.15256    \\              
 $f_{2}(1810)          $  & 1.72923     & $-$0.06709  & 0.22941   &  $-$0.43133   &  0.23721    \\              
 $f_{2}(1950)          $  & 1.96700     & $-$0.49924  & 0.19295  & 0.27975  &   $-$0.03987    \\              
 \hline 
	\end{tabular} \\
	\begin {tabular}{cccccccc}
    \noalign{\smallskip}
  \multicolumn{2}{c}{background terms} \\
	   $\tilde{c}_{00}$ & $-$0.04319 &&&&&&\\ 
	   $\tilde{c}_{01}$ &  0 & $\tilde{c}_{11}$ & 0 &  &&&\\
	   $\tilde{c}_{02}$ &  0.00984 & $\tilde{c}_{12}$ & 0 & $\tilde{c}_{22}$ & $-$0.07344 & &\\
	   $\tilde{c}_{03}$ &  0.01028 & $\tilde{c}_{13}$ & 0 & $\tilde{c}_{23}$ &  0.05533 & $\tilde{c}_{33}$ & $-$0.05183\\ 
	\hline\hline
	\end {tabular}
      \end{table}
      
  \begin{table}[h!]
     \caption{K-matrix parameters for the $\rho$-wave. Only constant
       background terms are taken into account. Indices 0
       to 2 stand for
       the channels $\pi\pi,2\pi\,2\pi$ and $K\bar{K}$. For the Chew-Mandelstam
       function, the $2\pi\,2\pi$ channel is treated as an effective
       $2\pi\,2\pi$ channel as described in~\cite{Albrecht:2019ssa}. The $\pi$-mass
       has been set to $m_{\pi^0}$, the $K$-mass to $m_{K^\pm}$ and
       the $\bar{K}$-mass to $m_{\bar{K}^0}$.}
     \label{tab:kmatrixrho0}
  \centering
	\begin {tabular}{cccccc} \\
	\hline\hline
	pole name $\alpha$&  $m^{\text{bare}}_{\alpha}$ & $g^{\text{bare}}_{\alpha \; \pi\pi}$ & $g^{\text{bare}}_{\alpha \; 2\pi\,2\pi}$ & $g^{\text{bare}}_{\alpha \; K\bar{K}}$ & \\ 
 $\rho(770)            $  & 0.71093     & 0.28023   & 0.01806  & 0.06501   &   \\     
 $\rho(1700)           $  & 1.58660     & 0.16318   & 0.53879  & 0.00495   &   \\     
 \hline
    \noalign{\smallskip}
  \multicolumn{2}{c}{background terms} \\
$\tilde{c}_{00}$ & $-$0.06948 & & & &   \\
$\tilde{c}_{01}$ &  0 & $\tilde{c}_{11}$ & 0 & &   \\
$\tilde{c}_{02}$ &  0.07958 & $\tilde{c}_{12}$ & 0 & $\tilde{c}_{22}$ & $-$0.60000   \\  
	\hline\hline
	\end {tabular}
   \end{table}

  \begin{table}[h!]
     \caption{K-matrix parameters for the $a_{0}$-wave. No
       background terms are taken into account. Indices 0
       to 1 stand for
       the channels $\pi\eta$ and $K\bar{K}$. The $\pi$-mass
       has been set to $m_{\pi^0}$, the $K$-mass to $m_{K^\pm}$ and
       the $\bar{K}$-mass to $m_{\bar{K}^0}$.}
     \label{tab:kmatrixrho0}
  \centering
	\begin {tabular}{cccc} \\
	\hline\hline
	pole name $\alpha$ &  $m^{\text{bare}}_{\alpha}$ & $g^{\text{bare}}_{\alpha \; \pi\eta}$ &
                                                                  $g^{\text{bare}}_{\alpha
                                                                  \; K\bar{K}}$ \\ 
 $a_0(980)            $  & 0.95395     & 0.43215   &  $-$0.28825  \\     
 $a_0(1450)           $  & 1.26767     & 0.19000   &   0.43372  \\     
 \hline \hline
   \end{tabular}
   \end{table}

  \begin{table}[h!]
     \caption{K-matrix parameters for the $a_{2}$-wave. Constant
       background terms are taken into account. Indices 0
       to 2 stand for the
       channels $\pi\eta,K\bar{K}$ and $\pi\eta'$. Note, that only these three decay channels were considered and other possibly significant channels are not included in our description. The $\pi$-mass
       has been set to $m_{\pi^0}$, the $K$-mass to $m_{K^\pm}$ and
       the $\bar{K}$-mass to $m_{\bar{K}^0}$.}
     \label{tab:kmatrixrho0}
  \centering
	\begin {tabular}{cccccc} \\
	\hline\hline
	  pole name $\alpha$ &  $m^{\text{bare}}_{\alpha}$ & $g^{\text{bare}}_{\alpha \; \pi\eta}$ & $g^{\text{bare}}_{\alpha \; K\bar{K}}$ & $g^{\text{bare}}_{\alpha \; \pi\eta'}$ &\\ 
 $a_2(1320)            $  & 1.30080     & 0.30073   &  0.21426 & $-$0.09162 & \\     
 $a_2(1700)            $  & 1.75351     & 0.68567   & 0.12543 & 0.00184 & \\
\hline
    \noalign{\smallskip}
  \multicolumn{2}{c}{background terms} \\
$\tilde{c}_{00}$ & $-$0.40184 & & & &   \\
$\tilde{c}_{01}$ &  0.00033 & $\tilde{c}_{11}$ & $-$0.21416 & &  \\
$\tilde{c}_{02}$ & $-$0.08707 & $\tilde{c}_{12}$ & $-$0.06193 & $\tilde{c}_{22}$ & $-$0.17435  \\  
	\hline\hline
   \end{tabular}
   \end{table}

 \begin{table}[h!]
     \caption{K-matrix parameters for the $\pi_1$-wave. Only constant
       background terms are taken into account. Indices 0
       to 1 stand for the
       channels $\pi\eta$ and $\pi\eta^\prime$.  Note, that only these two decay channels were considered and other possibly significant channels are not included in our description.  The $\pi$-mass
       has been set to $m_{\pi^0}$.}
     \label{tab:kmatrixrho0}
  \centering
	\begin {tabular}{cccc} \\
	\hline\hline
	pole name  $\alpha$ &  $m^{\text{bare}}_{\alpha}$ & $g^{\text{bare}}_{\alpha \; \pi\eta}$ & $g^{\text{bare}}_{\alpha \; \pi\eta^\prime}$ \\ 
 $\pi_1(1600)            $  & 1.38552     & 0.80564   & 1.04695      \\          
 \hline
    \noalign{\smallskip}
  \multicolumn{2}{c}{background terms} \\
$\tilde{c}_{00}$ & 1.0500 & &   \\
$\tilde{c}_{01}$ &  0.15163 & $\tilde{c}_{11}$ & $-$0.24611  \\  
	\hline\hline
	\end {tabular}
   \end{table}